\documentclass[journal]{IEEEtran}

\usepackage[utf8]{inputenc}
\usepackage{graphicx}
\usepackage{xcolor}
\usepackage{filecontentsdef}
\usepackage{soul}

\title{Passive and Privacy-preserving Human Localization\\ via mmWave Access Points for Social Distancing}
\author{Francesco Devoti,~\IEEEmembership{Member,~IEEE,} Vincenzo Sciancalepore,~\IEEEmembership{Senior Member,~IEEE,}\\ Xavier Costa-Perez,~\IEEEmembership{Senior Member,~IEEE}
\thanks{F. Devoti and V. Sciancalepore are with NEC Laboratories Europe GmbH, 69115 Heidelberg, Germany (e-mails: \{francesco.devoti,vincenzo.sciancalepore\}@neclab.eu).}
\thanks{X. Costa-Perez is with i2CAT Foundation, NEC Laboratories Europe GmbH, Heidelberg, Germany, and ICREA, Barcelona, Spain. (e-mail: xavier.costa@ieee.org).}
\thanks{This work was supported by EU H2020 RISE-6G project (No. 101017011), MINECO/NG EU
(No. TSI-063000-2021-6) and the CERCA Programme.}
}

\begin{document}
\maketitle
\begin{abstract}
The pandemic outbreak has profoundly changed our life, especially our social habits and communication behaviors. While this dramatic shock has heavily impacted human interaction rules, novel localization techniques are emerging to help society in complying with new policies, such as \emph{social distancing}. Wireless sensing and machine learning are well suited to alleviate viruses propagation in a privacy-preserving manner. However, its wide deployment requires cost-effective installation and operational solutions.

In public environments, individual localization information---such as social distancing---needs to be monitored to avoid safety threats when not properly observed.
To this end, the high penetration of wireless devices can be exploited to continuously \emph{analyze-and-learn} the propagation environment, thereby \emph{passively} detecting breaches and triggering alerts if required. In this paper, we describe a novel passive and privacy-preserving human localization solution that relies on the directive transmission properties of mmWave communications to monitor social distancing and notify people in the area in case of violations. Thus, addressing the social distancing challenge in a privacy-preserving and cost-efficient manner. Our solution provides an overall accuracy of about $99\%$ in the tested scenarios.
\end{abstract}

\section{Introduction}
\label{sec:intro}

The Covid-19 pandemic has turned everyone's lives upside down forcing national governments around the world to take measures to drastically reduce the rate of contagion. Localization information may support keeping \emph{social distancing} as well as \emph{contact tracing} of infected people thereby, playing a fundamental role in the fight against the virus spread.
The scientific community proposed \emph{social distancing} as the first effective measure against the uncontrolled virus spread, able to stop the transmission chains of the virus and prevent new ones from appearing.
This has resulted in strict rules to limit personal contacts and maintain interpersonal distance accordingly. However, in many situations (offices, grocery stores, shops, etc.) guaranteeing a social distance may be challenging as people can't continuously measure the inter-personal distance, especially if placed in small rooms or indoor environments, and some people might be less careful than others to comply social distancing.

This has potentially raised the industrial interest for developing advanced solutions to notify with alert messages upon minimum distance violation. They are mainly designed as infrastructure-free solutions that rely on wearable sensors based on ultra wide-band technology (UWB) or smartphones provided with bluetooth (BLE) or near-field communication (NFC) means, which discriminate whether the inter-social distance is below a certain threshold---by means of power measurements---thereby notifying the neighbors about the potential threat.
In addition, identity and contact duration information might result in privacy regulations violations~\cite{chan2021privacy}
The main drawback of such solutions is that the ability to detect interpersonal distance is only limited to people with such specific devices (or running given applications). This assumption might not hold in many cases, as neither telco operators nor national governments can force people in general to wear such devices or install such applications~\cite{nature_covid19}.

\begin{figure}
    \centering
    \includegraphics[width=\linewidth]{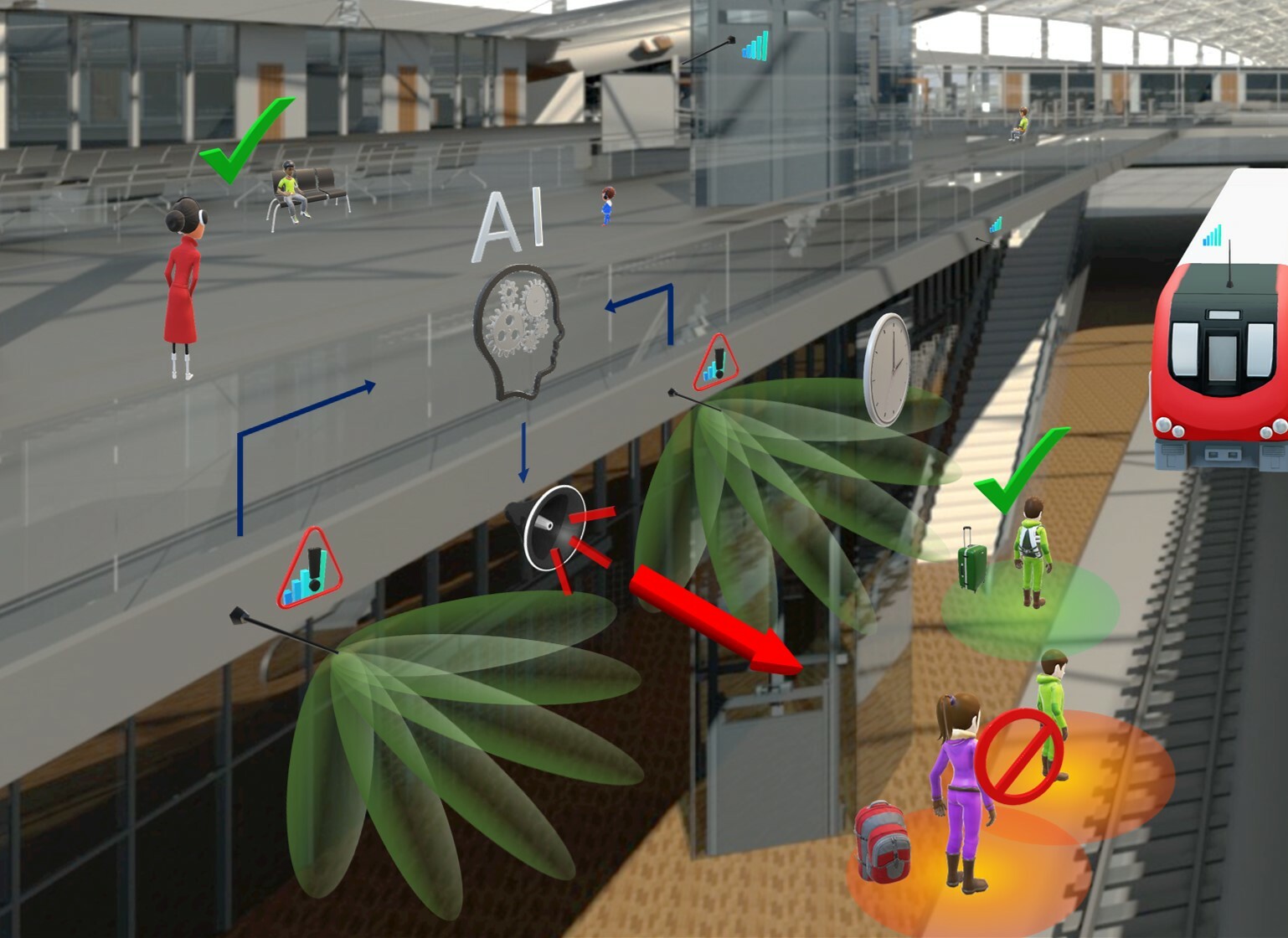}
    \caption{Passive detection of social distance in public scenarios using mmWave off-the-shelf devices provided with AI-based algorithms.}
    \label{fig:social_distancing_issue}
\end{figure}

Therefore, \emph{pervasive wireless passive} solutions shall be preferred, which gather localization information to detect undesired contacts and trigger exposure alerts in a privacy-preserving manner. These information can be used for many other use-cases, including intrusion detection, privacy-preserving health application, location-based service, etc. The development of 
advanced wireless sensing techniques are taking a prominent role thus in this space, since variations in wireless propagation signals can be exploited for such novel services.
Nonetheless, such applications generally require detailed and instantaneous information on the channel status, thus calling for specific radio-frequency hardware that can directly expose it via programmable interfaces~\cite{mec_covid19}. When only incomplete or partially hidden data is available, artificial intelligence (AI) might come to help: designed neural networks can continuously study the obtained pairwise channel information, learn unexpected variations and proactively map unusual fluctuations onto deterministic system state changes, i.e., the boolean state of social distancing violations in open and closed spaces.

In this work, we leverage the narrow directivity of millimeter-wave (mmWave) wireless communications to pioneer a \emph{new passive AI-based sensing system} capable of instantaneously detecting human gatherings violating social distancing policies and sending corresponding alerts to warn people. The proposed system does not rely on dedicated wearable devices, applications or active connections, thus it overcomes the limitations of infrastructure-free solutions.
At the same time, the system leverages reinforcement learning to automatically learn the human gathering detection task based on the observation of the reactions of people upon receiving the alert.
Hence, our passive solution is privacy-guaranteed and does not require human intervention.
To validate our proposed framework we carried out a synthetic trace simulations campaign and realistic deployment-based experiments with commercial mmWave devices in selected environments.
The proposed system can be seamlessly installed on existing WiFi network deployments in public or hybrid environments, such as rail stations (as depicted in Fig.~\ref{fig:social_distancing_issue}), airport terminals, bus stops, etc. with limited installation and maintenance costs.

\section{The technology race to limiting human interaction}
\label{sec:methods}

Localization information-based applications, such as social contact tracing, have become a key objective to effectively keep under control unwanted human virus spreading. However, this has proven to be a daunting task due to simultaneous lack of {\bf c1)} high reliability, {\bf c2)} extreme accuracy, {\bf c3)} agile deployment, and {\bf c4)} sustainable costs. We detail in the following the main effort in the literature towards those essential challenges, mostly taken individually.  

Computer vision and image processing from surveillance or dedicated cameras offer means to constantly monitor social distancing enforcement in public environments. On this line, \cite{AHMED2020102571} exploits the deep learning technique to efficiently carry out the object recognition process that can identify (and automatically locate) humans in video sequences and, in turn, estimate the euclidean distance among them. This is approximately performed by counting the pixels within the snapshots where people are detected. Similarly, \cite{SHORFUZZAMAN2021102582} proposes a deep learning-based framework that analyzes the data from mass video surveillance to monitor social distancing and trigger, in case, instantaneous alerts.
Such solutions are proven to be very effective in terms of high reliability and extreme accuracy ({\bf c1, c2}) but they require the installation of appropriate cameras, which might not be allowed in many everyday environments (e.g., offices, factories, hospitals, schools) or often might even incur in untenable costs.

In parallel, the very limited energy consumption of small (and sometimes wearable) devices enables to focus on infrastructure-free approaches capable of building low-cost and flexible ad-hoc networks ({\bf c3, c4}). Attaining such desirable objectives precludes relying on a supporting (and fixed) infrastructure while pushing for widespread technologies, such as bluetooth (BT) with low-energy capabilities (BLE), near-field communication (NFC) or ultra wide-band (UWB) sensors. In particular, \cite{nguyen2020comprehensive} provides specific means to combat the virus spread by minimizing the human approaching time, e.g. with ultrasound-based proximity detection systems or by means of wearable magnetic field proximity sensors.
Finally, \cite{covid19_mobicom20} blends together the need for accurate group tracking models with the infrastructure-free requirement that helps to detect contagion-related misbehavior within whatever environment. On top of dedicated hardware, tracing applications may leverage on complex wearable devices, such as smartphones, tablets, or smart watches, to enquiry public repositories thus notifying all direct human contacts about potential (diffusion) threat~\cite{covid19_infocom21}.

Most of the data thus collected shall be carefully (and efficiently) analyzed to rapidly block uncontrolled diffusion ({\bf c1, c4}).
This would require proper and complex mathematical models~\cite{TNSE_covid19} ({\bf c2}) that present scalability issues for very-crowded scenarios.
To cope with the complexity of the scenario, machine learning can be exploited to facilitate good approximations or to reconstruct hidden or unavailable data~\cite{eboost_jiot20}.

All above-mentioned techniques may help covering all presented challenges and thus preventing infections, only if combined together. Generally, this does not apply in realistic contexts, therefore pushing for a sustainable solution that can trade off all described features: our pioneering proposal paves the road towards an agile and flexible framework that can be readily installed on existing wireless infrastructures without requiring people to wear electronic devices but, at the same time, providing comparable accuracy and reliability levels of an infrastructure-free system.

\section{mmWave Directivity Gain to passively detect Environmental Change}
\label{sec:system_design}

While existing solutions might achieve reliability, fast actuation, or high responsiveness, exclusively, they can be fatally impaired by human misbehaviors. Therefore, there is an impelling need of an agile and flexible solution able to pursue high accuracy with affordable installation costs. 

Hereafter, we detail our solution that relies on conventional mmWave communication by shedding the light on the mathematical models and implementation details.

\subsection{Continuous mmWave 802.11ay channel monitoring}
Social distancing breaches can be ideally spotted by continuously monitoring the surrounding propagation environment to promptly detect suspicious variations. This operation can be performed in a passive way, wherein a transmitter pair interacts and keeps track of the channel response. Specifically, after establishing the mmWave link, power measurements can be regularly collected and analyzed to detect unexpected changes. To be compliant with IEEE 802.11ay standard guidelines\footnote{Details on the IEEE 802.11ay standard can be found at https://standards.ieee.org/standard/802\_11ay-2021.html}, power measurements are regularly performed during the beam training phase, i.e., when two mmWave devices discover each other by selecting the transmitting beam (direction) based on best response channel quality~\cite{wigig_infocom21}.

IEEE 802.11ay (like its predecessor 802.11ad) covers many relevant aspects to establish and sustain a communication link between mmWave-enabled devices. To provide the required beamforming capabilities, such devices are equipped with electronically steerable antenna arrays controlled by predefined weights vectors that are included in the so-called \textit{codebook}. Each wave vector in the codebook corresponds to the activation of a specific transmitting/receiving beam pattern. Those beam patterns are designed to be directional and their choice is subject to the instantaneous channel condition: a beam adaptation process is executed to avoid (nomadic) obstacles and efficiently follow the channel variations so that the communication is never disrupted.

The beam pattern selection is performed by means of a complex \emph{beam forming training} phase, wherein devices activate sequentially all available beam patterns---as per their codebook---and correspondingly collect power measurements that are used to select the best transmitting direction. It is started during the initial connection establishment (i.e., device paring phase) and periodically repeated to avoid connection drops~\cite{steinmetzer2017compressive}. In parallel, the wide spatial diversity provided by all available beam patterns allows obtaining a complete snapshot of the propagation environment, as shown in Section~\ref{sec:evaluation}, inspecting the surrounding area and keeping track of potential state changes.

\begin{figure}
    \centering
    \includegraphics[width=\linewidth]{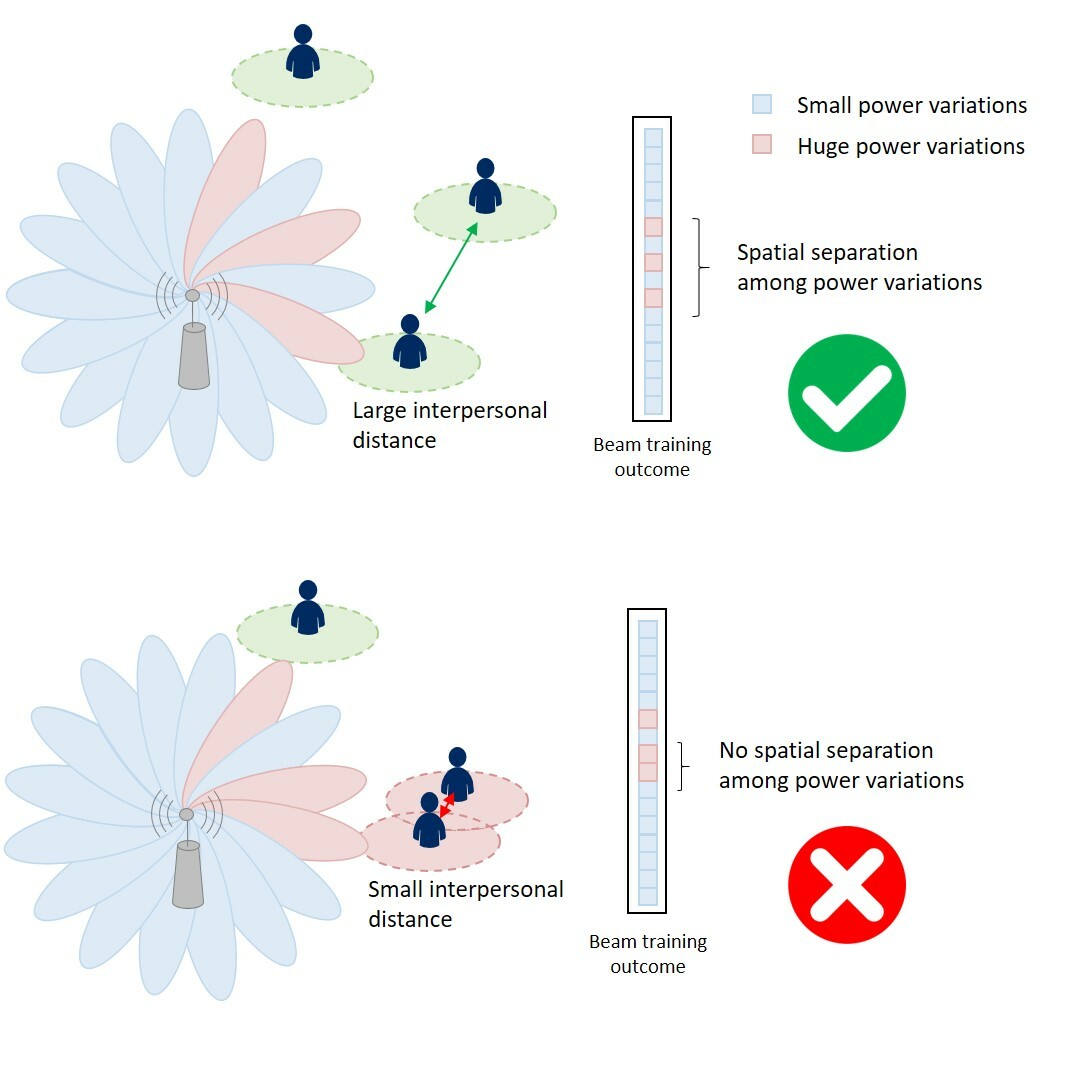}
    \caption{Effect of the interpersonal distance on the beam training outcome.}
    \label{fig:mm_w_benefits}
\end{figure}

\subsection{How to exploit the beam training phase to track interpersonal distance}
\label{sec:exploit_beam_training}
The high-frequency feature of mmWave communication hinders the overall signal propagation as it appears strongly affected by the propagation environment itself, including human bodies, walls, and even glass objects~\cite{slezak2018empirical}. On the one hand, this aspect may require an additional effort for properly designing a mmWave-based network to guarantee affordable communication quality levels. On the other hand, a passive environment monitor system may capitalize on this issue to build a reliable and low-cost solution that uses the effect on the short wavelength signal to continuously monitor selected areas.  
In particular, the different displacement of people in the environment can radically change the propagation conditions experienced by mmWave device pair.
Such changes are reflected in the outcome of the beam training procedure: we exploit such power measurements performed during standard operations to devise a complete sensing map of the propagation environment that can be smartly used to retrieve information on the environment itself, without requiring to extract advanced channel state information (CSI) from the devices or to deploy additional dedicated hardware. Hence, it will dramatically drop implementation costs.

Fig.~\ref{fig:mm_w_benefits} provides an example of different beam training phase outcomes as a function of the displacement of the people in the monitored area.
The example shows how the position of potential signal blockers (i.e., people) has an impact on the power measurements related to the different beam activation. Indeed, beams with directions towards potential signal blockers will experience larger attenuation with respect to the beams pointing towards non-blocked paths that translates into huge power variations reported in the beam training outcome. We leverage on this feature to build a system capable of detecting and reporting safe distance violations in areas covered by mmWave transmission service.
The rationale behind this is related to the beam training phase periodically performed between deployed devices (i.e., without involving user equipment), which we use to understand where blockages occur---based on selected transmitting/receiving beam and consequent directions---so as to infer the mutual distance between people, accordingly.
Naturally, as we show in Section~\ref{sec:evaluation}, the more directive the selected beam patterns, the higher the granularity of the environment sensing map and, consequently, the higher the sensitivity of the system against the position of the blocks, the higher the accuracy of the social distance detection model.

\section{AI-based mmWave channel monitoring}
\label{sec:ai_based_channel_monitorning}
Analytically modeling the effect of the interpersonal distance on the power measurements appears very challenging and biased due to the relevant dependence of the mmWave channel on the propagation environment. It is necessary to rely on machine learning techniques capable of automatically approximating the link between measured power and distance violation. This outstanding dependence on the environment turns into huge differences even in those areas that are relatively close to each other. Therefore, we analyzed a bench of classifiers specifically trained for each alert area.

\subsection{Safe distance violation detection system}
\label{sec:detection_system}

\begin{figure*}
    \centering
    \includegraphics[width=\textwidth]{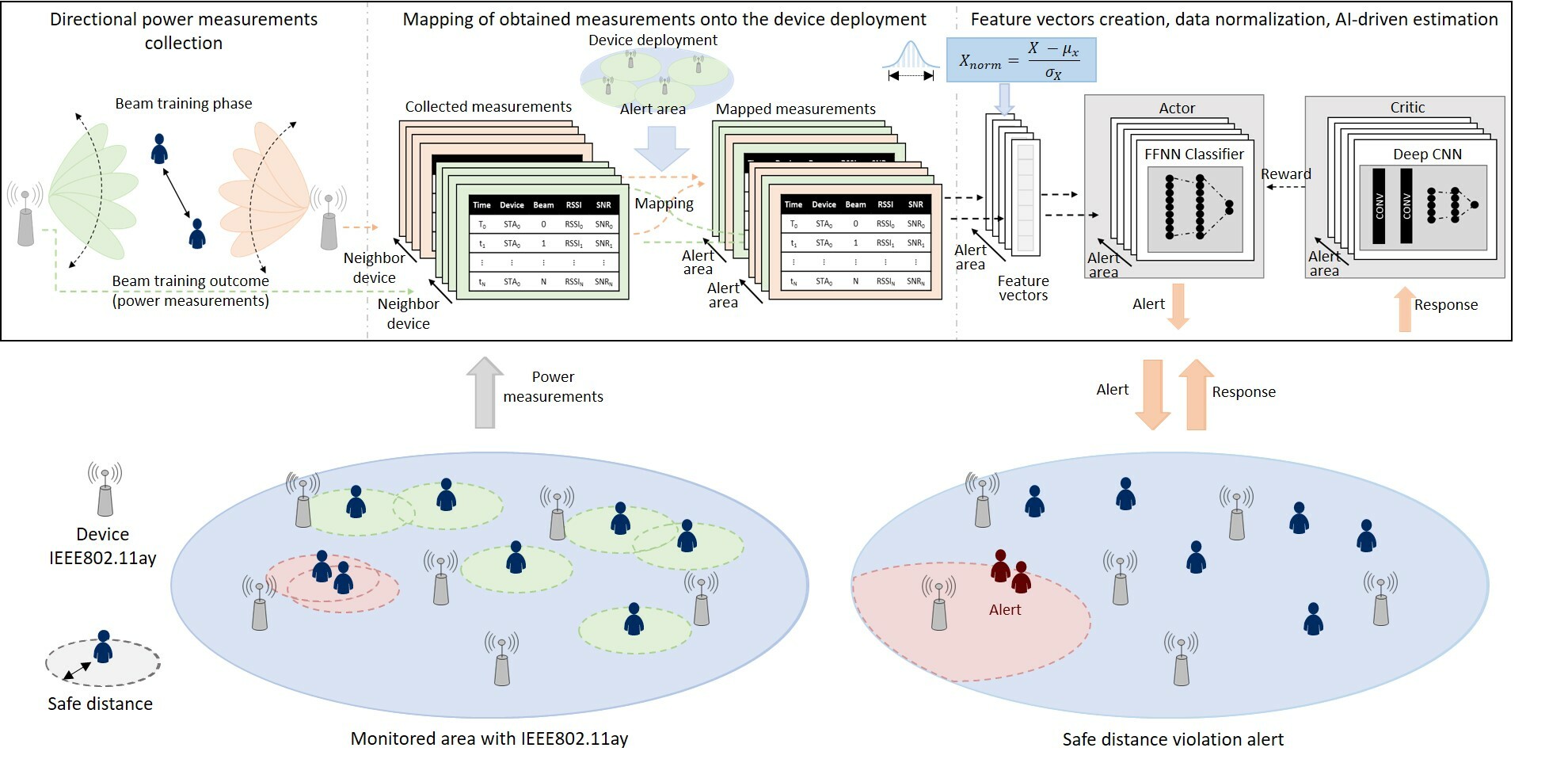}
    \caption{Safe distance violation detection and alert system. The system collects power measurements from device beam training phase, maps the obtained measurements onto the device deployment to define alert areas, and performs AI-driven human gathering estimation.  Training is performed via reinforcement-learning.}
    \label{fig:wireless_sensing}
\end{figure*}

Our safe distance violation detection and alert system is depicted in~Fig.~\ref{fig:wireless_sensing}. Let us consider a public area (e.g., train station, airport, shopping mall, office, etc.) wherein people, in accordance to the virus spread prevention measures, must keep a minimum safe distance, and wherein high-speed connectivity service is provided through IEEE 802.11ay-enabled access points (AP) (see Fig.~\ref{fig:social_distancing_issue}).

APs can provide connectivity to mobile stations (STAs) (e.g., smartphones, laptops, etc.), as well as fixed STAs (e.g., wireless displays, computers, cameras, etc.). Our system exploits the channel variations detected by mmWave devices. Those variations can be caused both by different displacement of blockers or movements of involved devices. Therefore, to filter out the device mobility effect, only devices that are fixedly deployed within the monitored area are considered as an information source for the system. Moreover, we assume the system to be deployed in controlled environments such as rail stations, shopping malls, etc., wherein the movements of nomadic obstacles (e.g., trains, buses, etc.) are periodically repeated in a quasi-deterministic fashion. Accordingly, their effect on the power measurements would affect many observations in the input of the classifier, which, given the high number of examples, would automatically filter out the contribution of such objects in the classification process.

According to the standard, devices in the area periodically activate the beam training procedure and perform power measurements. Note that APs and STAs can detect and collect beam training frames transmitted from nearby devices even if they are executing the connection handshake process.
The power measurements thus obtained are transmitted through a control plane to the safe distance violation system and collected to build a snapshot of the propagation environment state.
Given the limited coverage of mmWave access points, only a portion of the monitored area is considered relevant for the power measurement campaign related to a given pair of devices. Thus, we map all collected measurements onto the corresponding portion of the monitored area based on the device deployment and device coverage---assumed to be known---namely \textit{alert area}. This allows the system to send targeted alerts only to the specific areas wherein safe distance violations occur.
Note that we assume an optimal deployment phase being executed beforehand ensuring good channel conditions among devices to enable passive sensing: this allows to have strong communication paths, i.e., line-of-sight and/or second-order reflections, among devices covering a given alert area. Moreover, the denser the device deployment, the higher the granularity of the sensing/reflecting in smaller and more accurate target areas.

Such power measurements become the input feature of our detection system. For each alert area, the corresponding power measurements are collected and arranged into a feature vector whose elements contain power measurements corresponding to each activated beam pair and thus normalized. Feature vectors are fed into a bank of feed-forward neural network (FFNN) classifiers, one per alert area, which is in charge of performing safe distance violation detection, i.e., given the observation of the power measurements it provides the probability distribution over the set of classes \emph{safe distance violation}, and \emph{safe distance observance}.
We select FFNN due to their relatively low complexity and their ability of non-linear models generalization. However, different types of classifiers can be easily plugged into the proposed framework depending on the desired system efficiency and complexity.
It is worth pointing out that the lack of an active connection with the people in the area prevents our system from sending targeted notifications. Instead, if a violation is detected, a notification is sent to the specific alert area through advising systems, e.g., voice warnings or display boards. However, in situations where a group of people is not required to spread apart, e.g., people belonging to the same household, an unwanted alert might be triggered. Nonetheless, the alert is eventually sent to security officers in charge of evaluating the situation and enforcing safety policies within the target area if needed.

\begin{figure*}
    \centering
    \includegraphics[width=\textwidth]{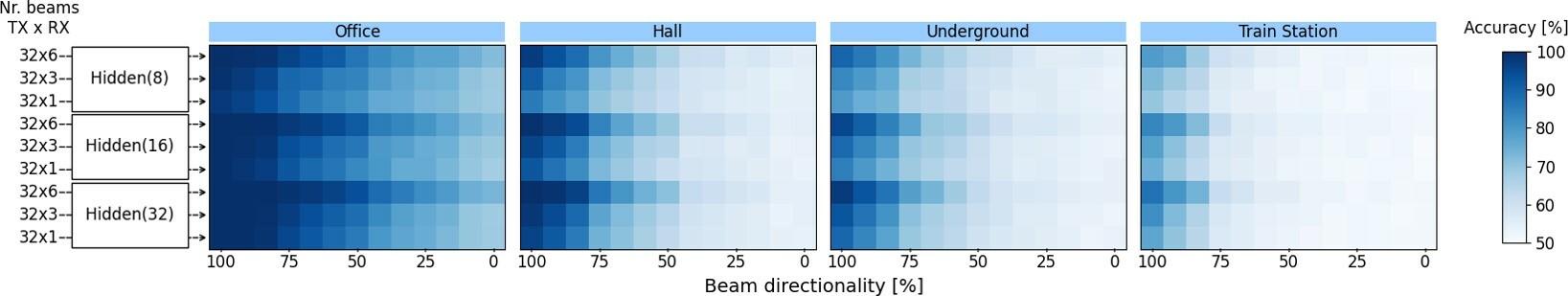}
    \caption{Accuracy of the distance violation detection process in four different simulated environments, considering different numbers of beams available at the devices $N$ and varying the beam directionality from a maximum of $100\%$ (half-power beamwidth equal to ${2\pi}/{N}$) to a minimum of $0\%$ (omnidirectional).}
    \label{fig:numerical_results}
\end{figure*}

\subsection{Reinforcement learning-based approach}
\label{sec:reinforcement_learning}

The bank of FFNN classifiers needs to be properly trained to detect safe distance violations.
To this extent, we rely on a reinforcement learning approach that exploits people's reaction to the alert to reveal the correctness of notified alerting messages and, based on this, it automatically learns how to truly detect safe distance violations.
The rationale behind this approach is the following. In the event of a safe distance violation detection, an alert is notified to the corresponding alert area. If the detection is correct, people in the area will automatically react to the alert by rearranging themselves to return in a safe condition (Hawthorne effect). This reaction reflects into a noticeable change of the propagation environment that can be clearly captured in the power measurement snapshots following the alert. 
Conversely, if the safe distance violation is incorrectly detected, the sent alert will not push people to move far away. Consequently, an imperceptible change in the snapshots following the alert will occur.

In the reinforcement learning architecture we propose, the actor network is in charge of classifying safe distance violations, while the critic is in charge of monitoring the channel variations following an alert and provide a reward to the actor, accordingly. For this reason, we design the critic as a deep CNN---which is fed with the series of feature vectors following the alert concatenated over time---and involving the convolutional layers, particularly suitable to recognize the space-time features characterizing the movement of people from its input.
This mechanism allows to automatically learn how to correctly detect safe distance violations independently from the environment in which the system is deployed.
It is worth highlighting that the learning mechanism we propose keeps unaltered the level of privacy guaranteed by our passive solution. Moreover, although relying on an instinctive reaction such as the Hawthorne effect, which in some cases could be ignored, alert messages surely reach people in charge of guarding the monitored area, thus forcing lawbreakers to return to a safe situation.

\section{Experimental evaluation}
\label{sec:evaluation}

Hereafter, we provide the performance evaluation of our system, which we carry out through a simulation campaign, where we generate synthetic beam training phases traces, as well as through a real implementation of our solution as a software module installed in four commercial IEEE802.11ad-enabled devices deployed within a real office environment.

\subsection{Synthetic scenario}
\label{sec:synthetic_scenario}
For the beam training synthetic traces generation, we emulate the system by means of an ad-hoc MATLAB\textsuperscript{\textregistered} simulator.
We consider four different experimental environments as follows:
\begin{itemize}
    \item \textit{office}, constituted by a $5m \times 5m$ squared indoor environment with $4$ mmWave-enabled devices deployed in the area, people can freely move within the area;
    \item \textit{hall}, constituted by a $10m \times 10m$ squared indoor environment with $4$ mmWave enabled devices deployed in the area, people can freely move within the area;
    \item \textit{underground}, constituted by a $10m \times 20m$ environment, wherein we place a $5m \times 20m$ platform. $4$ mmWave enabled devices are regularly deployed in the platform area, people can freely move on the platform;
    \item \textit{station}, constituted by a $20m \times 20m$ environment with two $5m \times 20m$ platforms separated from each other by a $10m$ space wherein train rails are located. Two mmWave enabled devices per platform are deployed, people can freely move on the platforms.
\end{itemize}

Millimeter-wave propagation and beam patterns are modeled as per~\cite{devoti2018mm}. Moreover, to take into account the typical non-idealities of beam patterns generated by commercial devices~\cite{steinmetzer2017compressive}, we vary the half-power width of the beams $W$ according to the equation $W=2\pi\left[1-\left(1-\frac{1}{N}\right)\alpha\right]$, where $N$ is equal to the number of transmitting/receiving beam configurations in the codebook, and $\alpha$ is a scaling factor that allows us to modulate the directionality in our experiments and ranging from a maximum value of $1$--- resulting in $\frac{2\pi}{N}$ wide beams corresponding to the $100\%$ of directionality---to a minimum value of $0$---resulting in a beam width of $2\pi$ corresponding to the $0\%$ of directionality, i.e., omnidirectional beam patterns.
We consider devices with a number of $32$ transmitting beams, and $1$, $3$, or $6$ receiving beams.
Following the IEEE 802.11ad standard, beam pattern alignment procedure is performed every beacon interval (i.e., every $10ms$) per each device pair in both transmitting and receiving directions.

Person bodies are emulated as fully absorbing cylinders with a radius of $0.25m$.
We randomly drop up to $6$ people in the simulation playground.
The minimum safe distance to be kept is set to $1.5m$ as usually recommended by European health care institutions to reduce the spread of the virus. We consider a total of $200$ different people arrangements in which social distancing regulations are violated, and additional $200$ people arrangements where the minimum social distance is fulfilled.
For each people displacement, we collect the power measurements of $200$ beam training procedures, for a total of $80000$ channel measurement snapshots that are arranged to form input feature vectors and normalized via \textit{independent standardization}. The overall dataset of snapshots is split according to a $60/20/20$ ratio for the purposes of training, validation, and testing procedures,
respectively.
The classification process is performed by a fully connected feed forward neural network (FFNN) with a single hidden layer of neurons with \textit{relu} activation function.
We train our neural network with a batch size of $1000$, a number of $30$ epochs, a learning rate of $0.001$ and \textit{Adam} optimizer.

Fig.~\ref{fig:numerical_results} shows the performance of our system in terms of safe distance violation detection accuracy achieved in the different scenarios we consider, with different beam pattern configurations, both in terms of number of transmitting/receiving beam patterns available as well as beam directionality. We consider different neural network complexity by varying the number of neurons forming the hidden layer.

From the results, it can be seen how the directional capabilities of the devices has a direct impact on the achievable performance. Indeed, the beam directionality directly affects the directional sensing effectiveness of the system, as the more directional the beams, the higher the granularity of the sensing map available to the system, thus, the better the achievable performance (the higher the system accuracy).
Additionally, the number of available beam patterns also affects the overall system performances. Indeed, increasing the number of beams increases the different points of view that the system can exploit to efficiently run the violation detection process, with a consequent performances increase.
On the AI algorithm, results show that different neural network complexities (i.e., number of neurons in the hidden layer) slightly change the system performance when a high number of beams with high directionality is available. While it has a greater impact when the devices are equipped with fewer beam patterns.
This is a direct consequence of the quality of the input features in relation, as previously described, to the devices directional capabilities. The better the quality of the input features, the easier the detection task and vice versa. This is reflected in the neural network complexity required to achieve high detection performances.

Finally, results show how the different considered scenarios impact on the system performances: the wider the monitored area, the lower the system performances. Recalling that we are keeping constant the number of deployed devices in the simulated alert areas, this behavior is mainly due to the density of devices involved in the measurement process, which naturally affects the system performance. Nonetheless, our system shows oustanding performances that vary from $~75\%$ to $~99\%$ depending on the selected scenario.

\begin{figure}
    \centering
    \includegraphics[width=0.8\linewidth]{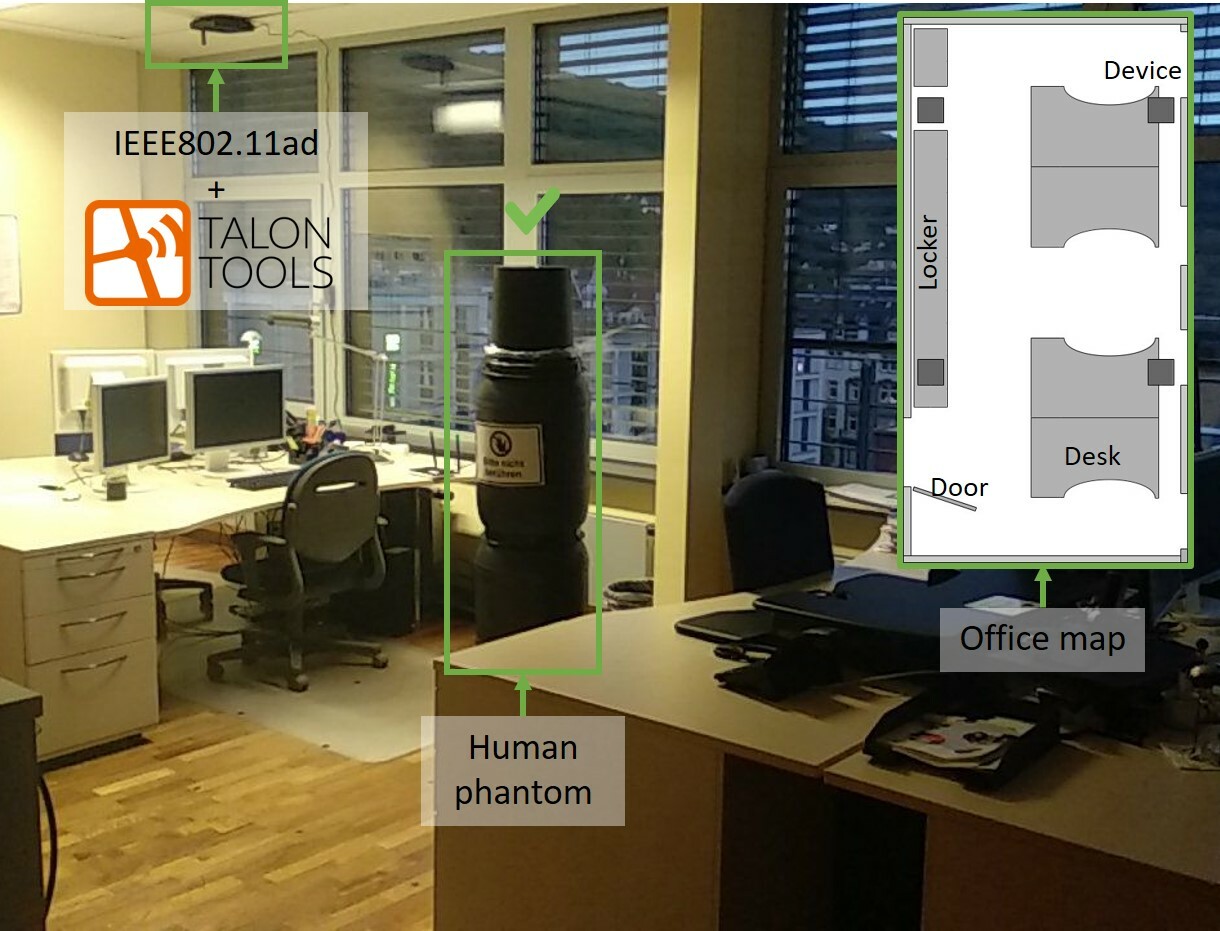}
    \caption{Executive experimental set in real office environment with four IEEE802.11ad devices mounted on the ceiling and human phantom. The proposed device deployment at the top right-hand corner, can efficiently and accurately cover the overall office environment.}
    \label{fig:testbed}
\end{figure}

\subsection{Real office environment}
\label{sec:real-scenario}
\begin{figure}
    \centering
    \includegraphics[width=0.9\linewidth ]{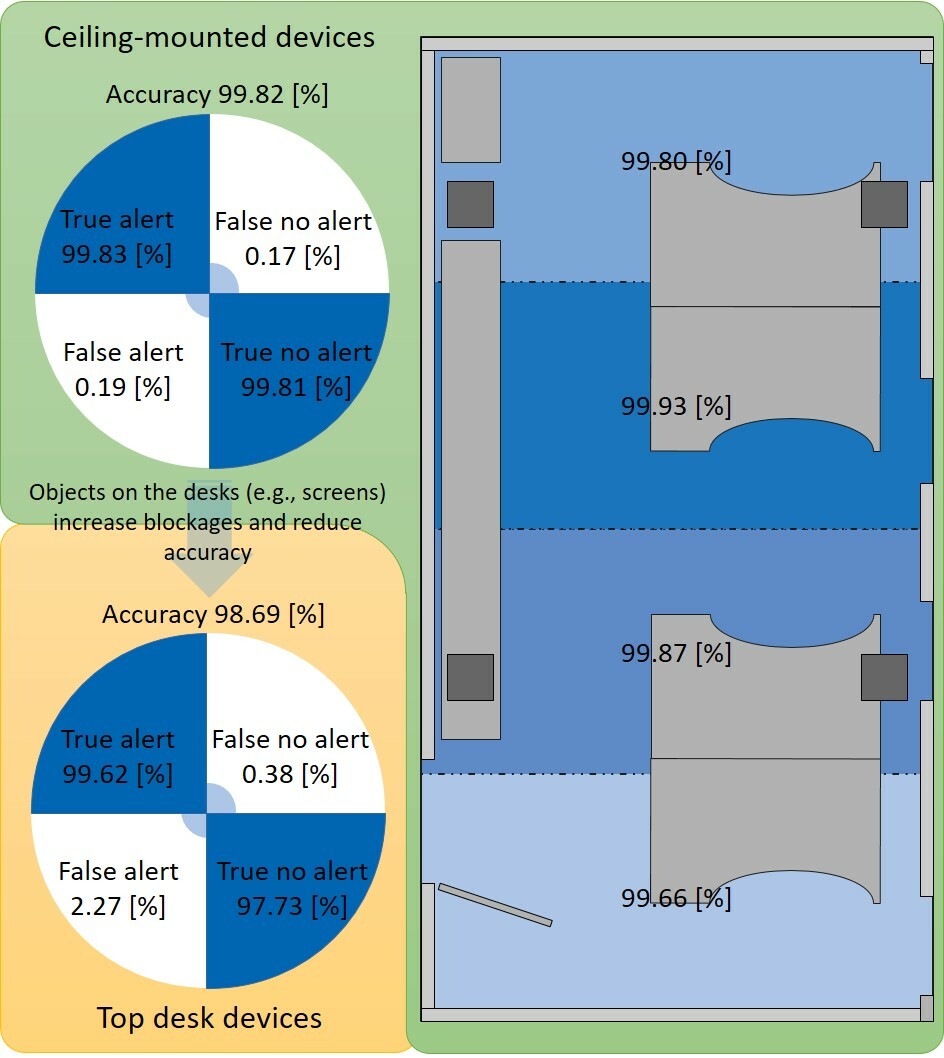}
    \caption{Accuracy of the distance violation detection process, obtained in a real office environment with two different configurations of the device deployment (left-hand side), and different areas of the office with the devices mounted on the ceiling (right-hand side).}
    \label{fig:real_scenario_results}
\end{figure}

To validate our system in a real environment, we implement our solution as a software module running on commercial off-the-shelf devices.
Thus considering realistic beamforming patterns and propagation conditions.
In particular, we deploy $4$ Talon ad7200 devices in a real office environment. Devices are provided with $36$ transmitting beams with different pointing directions and one quasi-omnidirectional receiving beam. The default device firmware does not include an easy access to the beam training and received power values, therefore we use the LEDE-ad7200 custom firmware~\cite{steinmetzer2017compressive} on such devices that allows us to retrieve the power measurements performed by devices during the beam training phase, and use them to reveal infractions of the minimum safe distance. To comply with current social distancing regulations, we emulated the presence of two people in the office with a real person and a human phantom. Fig.~\ref{fig:testbed} shows our testbed implementation, wherein devices are mounted on the ceiling according to the office map. Additionally, we consider a second deployment setup where the two devices on the right-hand side of the office map are placed on the top of the corresponding desk. We refer the reader to~\cite{pasid} for additional information on the testbed.

To validate our system, per each deployment setup of the devices, we consider a total of $40$ different dispositions of person and human phantom in the office: $20$ with safe distance violation and the remaining $20$ with sufficient interpersonal distance. For each disposition, we collect the beam training measurements performed by the devices for $3$ minutes. Thus obtaining a total of $2$ hours of measurements, with about $80000$ beam training procedures.

The measurements thus collected are divided according to a $60/20/20$ ratio for the purposes of training, testing and validation. We report in Fig.~\ref{fig:real_scenario_results} the performance obtained with a FFNN classifier with a single hidden layer constituted by $32$ neurons. We follow the same training process as described in Section~\ref{sec:synthetic_scenario}. 

From the obtained results it can be noticed that while our passive detection system can be easily installed on an existing WiFi network incurring in cost-effective, reliable and agile deployment, it is able to spot safe distance violations with an accuracy higher than $99\%$ when devices are mounted on the ceiling. Such performance is kept all over the office area, with a minimum variation depending on the relative proximity of devices and monitored persons. The overall accuracy is slightly lower when devices are placed according to the top desk setup, wherein office furniture (e.g., screens) causes higher attenuation between deployed devices, and makes channel variations caused by the presence of people more difficult to sense. Nevertheless, the overall accuracy in our settings is higher than $98\%$. However, in line with the simulations, accuracy performance might slightly reduce in other realistic deployments.

\section{conclusion}
\label{sec:conclusion}
The Covid-19 virus spreading explosion turned everyone's lives upside down. \emph{Social distancing} proved to be an effective measure to control virus spreading but unfortunately compliance was difficult for people due to deeply-rooted social habits.
In this paper, we have proposed an AI-based mmWave sensing solution that by passively monitoring changes on the wireless environment can infer localization information, thereby detecting social distancing breaches and triggering correction actions in a privacy-preserving manner without requiring an active connection with the user equipment.

Our advanced proposal combines cost-efficiency, agility, reliability, and accuracy challenges together into a novel passive detection system that can be supported by a variety of applications.
The solution has been evaluated through a simulation campaign and a real deployment with commercial mmWave devices. Proof-of-concept results show a promising detection accuracy of social distance above $99\%$. The system has been designed such that it can be seamlessly added as a software module to off-the-shelf commercial mmWave devices.

\bibliographystyle{IEEEtran}
\bibliography{references}

\begin{IEEEbiographynophoto}{Francesco Devoti} (M’20) received the B.S., and M.S. degrees in Telecommunication Engineering, and the Ph.D. degree in Information Technology from the Politecnico di Milano, in 2013, 2016, and 2020 respectively. He is currently a senior research scientist in the 6G Network group at NEC Laboratories Europe. His research interests include reflective intelligent surfaces, millimeter-wave technologies in 5G and 6G networks, and network slicing.
\end{IEEEbiographynophoto}

\begin{IEEEbiographynophoto}{Vincenzo Sciancalepore} (S'11--M'15--SM'19) received his M.Sc. degree in Telecommunications Engineering and Telematics Engineering in 2011 and 2012, respectively, whereas in 2015, he received a double Ph.D. degree. Currently, he is a senior 5G researcher at NEC Laboratories Europe, focusing his activity on network virtualization and network slicing challenges. He is currently the Industrial Chair of the IEEE Emerging Technologies Initiative on Reconfigurable Intelligent Surfaces. He is an Editor of the IEEE Transactions on Wireless Communications.
\end{IEEEbiographynophoto}

\begin{IEEEbiographynophoto}{Xavier Costa-Perez} (M'06--SM'18) is Head of Beyond 5G Networks R\&D at NEC Laboratories Europe, Scientific Director at the i2Cat R\&D Center and Research Professor at ICREA. His team contributes to products roadmap evolution as well as to European Commission R\&D collaborative projects and received several awards for successful technology transfers. Xavier served on the Program Committee of several conferences (including IEEE ICC and INFOCOM), published at top research venues and holds several patents. He received his M.Sc. and Ph.D. degrees in Telecommunications from the Polytechnic University of Catalonia (UPC) in Barcelona.
\end{IEEEbiographynophoto}
\vfill

\end{document}